\documentclass{elsarticle}

\journal{Physics Letters A}

\usepackage{amsfonts,amstext,graphics,graphicx,subfigure}
\usepackage[colorlinks]{hyperref}
\usepackage{color}
\def\XXint#1#2#3{{\setbox0=\hbox{$#1{#2#3}{\int}$ }
\vcenter{\hbox{$#2#3$ }}\kern-.5\wd0}}

\def\calc{\mathcal{C}}

\def\calh{\mathcal{H}}

\def\bq{\begin{equation}}
\def\eq{\end{equation}}
\def\bqy{\begin{eqnarray}}
\def\eqy{\end{eqnarray}}

\def\de{\delta}

\def\p{\partial}

\def\and{\quad\mathrm{and}\quad}

\def\cyc{\mathrm{cyc}}

\def\ncr{\nonumber\\}

\newcommand{\refeq}  [1] {(\ref{#1})}

\newcommand{\Half}{\frac{1}{2}}

\newcommand{\atX}[1]{\bigg|_{X_{#1}}}

\begin{document}

\title{Hamiltonian Nature of Monopole Dynamics}

\author[UT]{J.~M.~Heninger}
\ead{jeffrey.heninger@yahoo.com}

\author[UT]{P.~J.~Morrison}
\ead{morrison@physics.utexas.edu}

\address[UT]{Department of Physics and Institute for Fusion Studies, \\
The University of Texas at Austin, Austin, TX, 78712, USA}

%\author{J.~M.~Heninger}
%\email{jeffrey.heninger@yahoo.com}
%\affiliation{Department of Physics and Institute for Fusion Studies, 
%The University of Texas at Austin, Austin, TX, 78712, USA}
%\author{P.~J.~Morrison}
%\email{morrison@physics.utexas.edu}
%\affiliation{Department of Physics and Institute for Fusion Studies, 
%The University of Texas at Austin, Austin, TX, 78712, USA}

\date{\today}

\begin{abstract}
 
%\centering
%Electromagnetism \\
%with magnetic monopoles \\
%is not Hamiltonian.

Classical electromagnetism with magnetic monopoles is not 
a Hamiltonian field theory because the Jacobi identity 
for the Poisson bracket fails. 
The Jacobi identity is recovered only if all of the species 
have the same ratio of electric to magnetic charge or 
if an electron and a monopole can never collide. 
Without the Jacobi identity, there are no 
local canonical coordinates or Lagrangian action principle. 
To build a quantum field of magnetic monopoles, 
we either must explain why the positions of electrons and
monopoles can never coincide or 
we must resort to new quantization techniques.

% Key Words:  tba
\end{abstract}

\maketitle

%\tableofcontents

%%%%%%%%%%%%%%%%%%%%%%%%%%%%%%%%%
%%%%%%%%%%%%%%%%%%%%%%%%%%%%%%%%%

\section{Introduction}

	This letter considers the classical gauge-free theory of 
	electromagnetic fields interacting
	with electrically and magnetically charged matter 
	as a Hamiltonian field theory. 
	We begin with a brief history of magnetic monopoles 
	and then describe why monopole theories are not 
	Hamiltonian field theories.  

	The modern theory of magnetic monopoles was developed
	by Dirac \cite{Dirac31,Dirac48}. He showed that an electron
	in the magnetic field of a monopole is equivalent
	to an electron whose wave function is zero along a 
	semi-infinite `string' extending from the location of the 
	monopole. Along this string, the electromagnetic vector
	potential is undefined. The phase of the electron is no 
	longer single valued along a loop encircling the Dirac 
	string. In order for observables to be single valued, the 
	phase shift must be an integer multiple of $2\pi$, so
	the electric and magnetic charge must be 
	quantized. The direction of the string is arbitrary. 
	Changing it corresponds to a gauge transformation for the 
	fields and a global phase shift for the wave function. 
	To avoid the string entirely, we could instead
	define the vector potential for multiple patches around
	the monopole \cite{WuYa76}.
	
	Magnetic monopoles can also be introduced in the
	hydrodynamic formulation of non-relativistic quantum
	mechanics \cite{Bir71}. Since this formulation 
	involves fluid-like variables and the fields, it removes the 
	ambiguity associated with the wave function and vector 
	potential. Dirac strings are replaced by singular vortex
	lines. 
%	Evaluating the expectation value of any observable
%	involves a line integral along a path which does not 
%	pass through any of the singular vortex lines.

	Quantum field theories for magnetic monopoles and for
	dyons (particles with both electric and magnetic charge)
	were developed by Cabibbo \& Ferrari \cite{CabFer62} and
	Schwinger \cite{Schw66,Schw68,Schw69}. These theories use
	two nonsingular vector potentials that are related to the
	fields by a convolution with a string function. This
	string function allows for the derivation of the action
	and equations of motion for interacting electron and
	monopole fields.
	
	Grand unified theories (GUTs) describe the strong, weak, 
	and electromagnetic forces as a single theory whose 
	symmetry is spontaneously broken at lower energies. 
	If the symmetry is not broken in the same direction 
	everywhere, then the fields will be zero at some locations.
	Around these locations, the fields resemble the fields 
	of a magnetic monopole \cite{tHo74,Poly74}.
	Magnetic monopoles are a generic feature of GUTs, including
	string theories \cite{Pol04}.
	
	If we assume that a GUT exists, then, shortly after
	the Big Bang, the expanding universe cooled through the 
	critical temperature at which the symmetry is broken.
	There is no reason to assume that the symmetry would be
	broken in the same direction at causally disconnected
	locations. The boundaries between regions with differently 
	broken symmetry would produce magnetic monopoles and 
	strings \cite{Kib76}. Initial estimates of the number 
	of monopoles produced this way were much too
	high \cite{Pres79}, but the estimates are dramatically 
	reduced by inflation \cite{Guth81}.
	
	The existence of astronomical magnetic fields produces
	a bound on the number of monopoles. If there were too many
	monopoles, they could move and screen out large magnetic 
	fields, much like electrically charged matter screens out
	large electric fields in our universe \cite{Par70}.
	
	Direct observations of magnetic monopoles remain 
	inconclusive. Two early experiments detected candidate
	events \cite{Pri75,Cab82}, but one was immediately
	refuted \cite{Alv75} and the other has never been replicated.
	Extensive searches for monopoles have been done
	in matter, in cosmic rays, via catalyzing nucleon decays,
	and at colliders, all with negative results \cite{Pat16}.
	
	More information about magnetic monopoles can be found
	in one of the many relevant review articles
	\cite{Mil06,Raj12} or textbooks \cite{Atiyah,Shnir}.
	
	The letter is organized as follows. 
	In Sec.~\ref{sec:VM}, we introduce a general and 
	standard matter model for plasmas, 
	the Vlasov-Maxwell equations, 
	as a noncanonical Hamiltonian field theory, 
	i.e.\ one without the standard Poisson bracket. 
	We add monopoles to this theory in Sec.~\ref{sec:VMM}
	and show that it no longer satisfies the Jacobi identity, 
	a basic premise of Hamiltonian theory. 
	Sec.~\ref{sec:1E1M} considers whether the interaction between
	a single electron and a single magnetic monopole is 
	Hamiltonian. We discuss the importance of the Jacobi
	identity in Sec.~\ref{sec:Jac} and consider the difficulties
	in quantization without the Jacobi identity in 
	Sec.~\ref{sec:Quan}. We conclude in Sec.~\ref{sec:Con}.
	
\section{Vlasov-Maxwell Equations}
\label{sec:VM}

	We approach the problem of magnetic monopoles from the 
	perspective of plasma physics, although our
	conclusion is general and independent 
	of the Vlasov-Maxwell matter model. 
	In plasmas, the most important 
	dynamics are the collective motions of the
	particles in collectively generated electromagnetic fields.
	The relevant dynamical variables are 
%	not the individual
%	particles' degrees of freedom, but rather 
	the distribution function $f_s(x,v,t)$
	for each species $s$, which describes the 
	probability density of the particles in phase space,
%	of finding a particle of species $s$ with 
%	position between $x$ and $x+dx$ and velocity between 
%	$v$ and $v+dv$, 
	and the electric and magnetic fields $E(x,t)$ and $B(x,t)$. 
	The charge and mass of each species	are $e_s$ and $m_s$. 
%	If we integrate $f_s$ over all positions and velocities, 
%	we get the total number of particles of that species, $N_s$.
	The temperatures of plasmas are high enough and the 
	densities are low enough that quantum effects are
	negligible. 
	
	The dynamics of the distribution function are governed by
	a mean-field transport equation. The phase space density is constant
	along particle trajectories: 
	\bq
		\frac{d f_s}{dt} = \frac{\p f_s}{\p t} + 
		v \cdot \frac{\p f_s}{\p x} + \frac{e_s}{m_s} 
		\left(E + \frac{v}{c} \times B\right) \cdot 
		\frac{\p f_s}{\p v} = 0 \,. \label{VlaMax}
	\eq
	This is combined with Maxwell's equations for the 
	electric and magnetic fields, with sources determined by
	the moments of the distribution function, 
	\bq
		\rho = \sum_s e_s \int f_s \,dv \,, \quad
		j = \sum_s e_s \int f_s \,v \,dv \,.
	\eq
	The resulting Maxwell-Vlasov equations are a closed system 
	of nonlinear partial integro-differential equations for 
	$f_s$, $E$, and $B$. Many reductions have been developed
	to convert these equations to more manageable forms
	like gyrokinetics and magnetohydrodynamics. 
	Since the Maxwell-Vlasov equations are more general than
	fluid equations, these results are generic 
	for matter models without dissipation.
	
%	Two particle interactions can be introduced by adding a 
%	collision operator, which scatters particles in velocity, to
%	the right hand side of \refeq{VlaMax}. 
%	Alternatively, the exact dynamics of a finite number of
%	interacting particles 
%	can be recovered by writing the distribution function as
%	a sum of Dirac delta functions in position and velocity.
	
	In 1931, Dirac wrote that ``if we wish to put the equations
	of motion [of electromagnetism] in the Hamiltonian form,
	however, we have to introduce the electromagnetic potentials'' \cite{Dirac31}.
	Born and Infeld showed that this is not entirely true
	\cite{Born1,Born2}.   
	The Vlasov  description of matter  coupled with Maxwell's equations can 
	be written as a Hamiltonian theory without introducing 
	potentials if we allow for a noncanonical Poisson bracket 
	(see Refs.~\cite{pjm80,pjm82,MW82,Bir84,Morr98} 
	for review). 
	The Hamiltonian functional and noncanonical Poisson bracket 
%	(see Refs.~[\onlinecite{pjm80}, \onlinecite{pjm82}, 
%	\onlinecite{MW},\onlinecite{Bir84}]) 
	for this system are
	\bq
		\calh = \sum_s \frac{m_s}{2}\!\! \int \!|v|^2 f_s \,d^3 x \,d^3 v + 
		\frac{1}{8 \pi}\!\! \int \!(|E|^2 + |B|^2) \,d^3 x \,,
		\label{VMHam}
	\eq
	\bqy
		\{F,G\} &=& \sum_s \int \Big(\frac{1}{m_s} \,f_s \left(\nabla 
		F_{f_s} \cdot \p_v G_{f_s} - \nabla G_{f_s}	\cdot \p_v F_{f_s}
		\right) \ncr
		&+ &  \,\frac{e_s}{m_s^2 c} \,f_s \,B \cdot \left(\p_v F_{f_s} \times
		\p_v G_{f_s} \right) \ncr
		&+ &  \,\frac{4\pi e_s}{m_s} \,f_s \left(G_E \cdot \p_v F_{f_s} -  
		F_E \cdot \p_v G_{f_s} \right) \Big) d^3 x \,d^3 v \ncr
		& +&  \,4 \pi c \int \left( F_E \cdot \nabla \times G_B - 
		G_E \cdot \nabla \times F_B \right) d^3 x  \,, 
		\label{VMPB}
	\eqy
	where subscripts such as $F_{f_s}$ refer to the functional
	derivative of $F$ with respect to $f_s$.
	
	The Hamiltonian, but not the Poisson bracket,
	needs to be modified to make this theory relativistic.
	Replace the $|v|^2$ in the kinetic energy term with 
	$\gamma |v|^2 = c^2 \sqrt{1+|v|^2/c^2}$ 
	(e.g.\ \cite{Bir84}).
	
	It is straightforward to derive the Vlasov equation 
	\refeq{VlaMax} and the dynamical Maxwell equations by 
	setting 
	\bq
		\frac{\p f_s}{\p t} = \{f_s, \calh\} \,, \,
		\frac{\p E}{\p t} = \{E, \calh\} \,, \,
		\frac{\p B}{\p t} = \{B, \calh\} \,.
	\eq
	The constraints appear as Casimir invariants:
	\bqy
		\calc_E &=& \int h_E(x) \left(\nabla \cdot E - 
		4 \pi \sum_s e_s \int f_s \,d^3 v \right) d^3 x \,, \\
		\calc_B &=& \int h_B(x) \,\nabla \cdot B \, d^3 x \,,
	\eqy 
	where  $h_E(x)$ and $h_B(x)$ are arbitrary functions.  The Poisson bracket of 
	$\calc_E$ or $\calc_B$ with anything
	is zero. Since the time dependence of anything is determined
	by its bracket with the Hamiltonian, Casimirs are 
	conserved for any Hamiltonian. If the Casimirs are initially
	zero (as required by the divergence Maxwell's Equations),
	they will remain zero for all time.
	
	There is an important subtlety to this formulation of 
	electromagnetism. If a system is Hamiltonian, its Poisson
	bracket must satisfy the Jacobi identity for any
	functionals $F$, $G$, and $H$:
	\bq
		\{\{F,G\},H\} + \{\{G,H\},F\} + \{\{H,F\},G\} = 0 \,.
		\label{Jacobi}
	\eq
	For the Vlasov-Maxwell system it was shown by direct calculation in \cite{pjm82, Morr13} that 
	\bqy
		\{F,\{G,H\}\} + \cyc &=&
		\label{DivBPB}\\
		&&\hspace{-2.5cm}
		\sum_s \int f_s \,\nabla \cdot B \,\left(
		\p_v F_{f_s} \times \p_v G_{f_s} \right) \cdot 
		\p_v H_{f_s} d^3 x \, d^3 v \,,
		\nonumber
	\eqy
	which means the domain of functionals must be restricted to solenoidal vector fields, $\nabla\cdot B=0$, 
	or equivalently defined on closed but not necessary exact two-forms.  Such a set of functionals is closed
	with respect to the bracket. 
%	because $\{F,G\}$ will also have this property. 
	
	The two Casimirs mentioned above are not symmetric.
	The value of $\calc_E$ could %, in principle,
	initially be chosen to be anything. If $\calc_B \neq 0$,
	i.e.\  if $\nabla \cdot B \neq 0$, then the Vlasov-Maxwell
	system would cease to be a Hamiltonian field theory.
	
	This is our first indication that the existence of 
	magnetic monopoles is connected to the Hamiltonian nature
	of classical electromagnetism.
	
\section{Vlasov-Maxwell with Monopoles}
\label{sec:VMM}

	What happens when we add magnetic monopoles? 
	
	We must change the Hamiltonian and/or the Poisson bracket
	so they produce the new equations of motion.
	
	The appropriate Hamiltonian, Poisson bracket, 
	and a detailed proof of the Jacobi identity
	were given in Section IV D and Appendix 3 of
	\cite{Morr13}.
%	Ref.~\onlinecite{Morr13}.
	For species $s$ with  electric charge $e_s$ and magnetic
	charge $g_s$ the Hamiltonian is identical to (\ref{VMHam}) and the Poisson bracket is 
	\bqy
		\{F,G\} &=& \sum_s \int \Big(\frac{1}{m_s} \,f_s \left(\nabla 
		F_{f_s} \cdot \p_v G_{f_s} - \nabla G_{f_s}	\cdot \p_v F_{f_s}
		\right) \ncr
		& +&  \,\frac{e_s}{m_s^2 c} \,f_s \,B \cdot \left(\p_v F_{f_s} \times
		\p_v G_{f_s} \right) \ncr
		& -&  \,\frac{g_s}{m_s^2 c} \,f_s \,E \cdot \left(\p_v F_{f_s} \times
		\p_v G_{f_s} \right)	
		\label{VMPBM}\\
		& +&  \,\frac{4\pi e_s}{m_s} \,f_s \left(G_E \cdot \p_v F_{f_s} -  
		F_E \cdot \p_v G_{f_s} \right) \ncr
		&+ &  \,\frac{4\pi g_s}{m_s} \,f_s \left(G_B \cdot \p_v F_{f_s} -  
		F_B \cdot \p_v G_{f_s} \right) \Big) d^3 x \,d^3 v \ncr
		& +& \,4 \pi c \int \left( F_E \cdot \nabla \times G_B - 
		G_E \cdot \nabla \times F_B \right) d^3 x \,.
		\nonumber
	\eqy
	The Jacobi identity for this bracket is
		\bqy
		\{F,\{G,H\}\} + \cyc &=& 
		\label{mjac}\\
		&& \hspace{-2cm}
		\sum_s \frac{1}{m_s^2}
		\int \p_v H_{f_s} \cdot 
		\left( \p_v F_{f_s} \times \p_v G_{f_s}\right) \ncr
		& &\hspace{-1cm} \times \,f_s
		 \left(e_s \nabla \cdot B - g_s \nabla \cdot E\right) 
		d^3 x \,d^3 v  \nonumber \,. 
	\eqy
	For (\ref{mjac}) to vanish for arbitrary $F$, $G$, $H$, and $f_s$, we must have
	\bq
		e_s \nabla \cdot B = g_s \nabla \cdot E \,, \quad
		\forall s \,.
	\eq
	
	The case when every species has the same ratio
	of magnetic to electric charge is addressed in 
	Section 6.11 of Jackson	\cite{Jackson}. 
	Using the duality transformation 
	\bqy
		E' = E \cos\xi + B \sin\xi &\,,\ &
		B' = -E \sin\xi + B \cos\xi \,, \\
		e'_s = e_s \cos\xi + g_s \sin\xi &\,,\ &
		g'_s = -e_s \sin\xi + g_s \cos\xi \,,
	\eqy
	with $\xi = \arctan(g_s/e_s)$, 
	the  magnetic charges are removed
	and the  Jacobi identity is satisfied.
	These monopoles are trivial; this theory is
	equivalent to electromagnetism without monopoles. 
	``The only meaningful question is whether \textit{all}
	particles have the same ratio of magnetic to electric 
	charge'' \cite{Jackson}.
	
	If not all species have the same ratio of magnetic to 
	electric charge, then the only way that the Jacobi identity 
	could be satisfied is if $\nabla \cdot E = 
	\nabla \cdot B = 0$. This is obviously not true in 
	general.
	
	When we add nontrivial magnetic monopoles to the 
	Vlasov-Maxwell system, the Jacobi identity is not 
	satisfied, so it is not a Hamiltonian field theory. 
	
\section{One Electron and One Monopole}
\label{sec:1E1M}

	Although we originally derived this result for the 
	collective motion of many electrically and magnetically
	charged particles, it should also hold when there are 
	only a small number of particles. 
	Consider the interaction between a single electron 
	with position $X_e$, velocity $V_e$, mass $m_e$, electric 
	charge $e$, and magnetic charge $0$ and a single monopole 
	with position $X_m$, velocity $V_m$, mass $m_m$, electric 
	charge $0$, and magnetic charge $g$.
	
	The Hamiltonian and Poisson bracket for this system follow
	from localizing on particles. Set
	\bq
		f_s= \de(x-X_{s}) \, \de(v-V_{s}) \,, \quad s = e,m \,.
	\label{deltaf}
	\eq 
	For the Poisson bracket, use the chain rule 
	expressions
	\bq
		\frac{\p  {F}}{\p X_s} = \left. 
		\nabla \frac{\de F}{\de f}\right|_{(X_s, V_s)}
	 	\mathrm{and}\quad
	 	\frac{\p  {F}}{\p V_s} =  \left.
	 	{\p_v } \frac{\de F}{\de f}\right|_{(X_s, V_s)}\,,
		\label{ptlchain}
	\eq
	where on the left of each expression, $F$ is the function of
	$(X_s, V_s)$ obtained upon substituting (\ref{deltaf}) into
	the functional $F$ on the right.
	This yields
	\bq
		\calh = \Half m_e V_e^2 + \Half m_m V_m^2 + 
		\frac{1}{8\pi} \int \left(|E|^2 + |B|^2\right) d^3 x \,,
	\eq
	\bqy
		\{F,G\} &=& \frac{1}{m_e} \left(\frac{\p F}{\p X_e} \cdot 
		\frac{\p G}{\p V_e} - \frac{\p G}{\p X_e} \cdot 
		\frac{\p F}{\p V_e}\right) \ncr
		& & + \frac{1}{m_m} \left(\frac{\p F}{\p X_m}
		\cdot \frac{\p G}{\p V_m} - \frac{\p G}{\p X_m} \cdot 
		\frac{\p F}{\p V_m}\right) \ncr
		& & + \,\frac{e}{m_e^2 c} \,B(X_e) \cdot \left(\frac{\p F}{\p V_e}
		\times \frac{\p G}{\p V_e} \right) \ncr
		& & - \frac{g}{m_m^2 c} \,E(X_m)
		\cdot \left(\frac{\p F}{\p V_m} \times \frac{\p G}{\p V_m} \right) \label{PB1E1M} \\
		& & + \,\frac{4\pi e}{m_e} \left(\frac{\de G}{\de E} \atX{e} \cdot
		\frac{\p F}{\p V_e} - \frac{\de F}{\de E} \atX{e} \cdot 
		\frac{\p G}{\p V_e}\right) \ncr
		& & + \frac{4\pi g}{m_m} 
		\left(\frac{\de G}{\de B} \atX{m} \cdot	\frac{\p F}{\p V_m} -
		\frac{\de F}{\de B} \atX{m} \cdot \frac{\p G}{\p V_m}\right) \ncr
		& & + \,4\pi c \int \left(\frac{\de F}{\de E} \cdot \nabla \times
		\frac{\de G}{\de B} - \frac{\de G}{\de E} \cdot \nabla \times
		\frac{\de F}{\de B} \right) d^3 x \,. \nonumber
	\eqy

	This Hamiltonian and Poisson bracket give the 
	expected equations of motion: the Lorentz force laws
	and the dynamical Maxwell equations, with currents
	proportional to $V_s \, \delta(x-X_s)$.
	The divergence Maxwell equations, with delta function 
	sources, appear in the Casimirs.
	
	The Jacobi identity calculation for (\ref{PB1E1M})
	can be done directly,  
	but it follows easily upon substituting (\ref{deltaf})
	and the second of (\ref{ptlchain})
	into (\ref{mjac}), yielding
	\bqy
		\{\{F,G\},H\}&+& \cyc =
        \frac{12 \pi e g}{c} \,\delta(X_e-X_m) \,
        \label{2jac}\\
        & \times&\Bigg(\frac{1}{m_e^3}
		\frac{\p F}{\p V_e} \cdot \left(\frac{\p G}{\p V_e}
		\times \frac{\p H}{\p V_e}\right) \ncr
		& & \hspace{.75cm} - \frac{1}{m_m^3} \frac{\p F}{\p V_m}
		\cdot \left(\frac{\p G}{\p V_m} \times 
		\frac{\p H}{\p V_m}\right) \Bigg) .	\nonumber
	\eqy
	The  Jacobi identity is not satisfied globally.
	There is a singularity when the positions of the electron
	and monopole coincide.
	
	Classically, 
	there is no reason why this coincidence couldn't happen.
	A stationary monopole produces a radial
	magnetic field. An electron moving directly towards the 
	monopole experiences a force $e V_e \times B / c = 0$.
	The electron passes through the monopole without 
	experiencing any force at all.
	
	This singularity is very different from the singularity
	for two electrically charged particles. That singularity
	comes from the Hamiltonian, can only be reached with
	infinite energy, and is removed if the point particles
	are replaced by continuous charge distributions. This
	singularity comes from the Jacobi identity, requires
	no energy to reach, and becomes worse if the point particles
	are replaced by continuous distributions because the Jacobi
	identity is violated at more locations.
		
	The electromagnetic interaction
	between a single electron and a single magnetic monopole 
	is not, in general, Hamiltonian.

\section{Importance of the Jacobi Identity}
\label{sec:Jac}

	Electromagnetism with magnetic monopoles does not satisfy
	the Jacobi identity. Why should we care? 
%	Are other theoretical arguments in favor of magnetic
%	monopoles strong enough to relax the requirement that
%	fundamental physical theories be Hamiltonian?
	
	There is extensive literature on the algebraic and geometric
	nature of Hamiltonian mechanics (e.g.\
	\cite{sudarshan,ArKozNei,souriau,thirring,DA15}) 
	with phase space defined as 
	a symplectic or Poisson manifold. 
	The Jacobi identity is central to these results. 
 
	Darboux's theorem, when applied to Hamiltonian systems, 
	says that the Jacobi identity implies the existence 
	of a local transformation to canonical coordinates on a
	foliation parameterized by Casimir invariants \cite{LJ82}. 
	For electromagnetism, this transformation occurs
	when we introduce the potentials: the Poisson bracket
	becomes simple and the Hamiltonian becomes more complicated. 
	
	If we apply an arbitrary coordinate	transformation to a
	bracket that satisfies the Jacobi identity, the new 
	bracket will also satisfy the Jacobi identity \cite{Eis61}.
	If the Jacobi identity is not satisfied, then no 
	coordinate transform can turn it into a canonical bracket.
	Local canonical coordinates do not exist.
	
	Most fundamental physical theories begin as a Lagrangian 
	action principle. If you have a Lagrangian action principle
	and the Legendre transform exists, then you can transform
	it into a Hamiltonian system with a Poisson bracket that
	satisfies the Jacobi identity. Contrapose this. If your
	system has a Poisson bracket that doesn't satisfy the Jacobi
	identity, then no Lagrangian action principle exists. 

\section{Quantizing without the Jacobi Identity}
\label{sec:Quan}

	Standard methods of quantization fail without the Jacobi 
	identity. Typically, we replace dynamical variables with
	operators whose commutation relation algebra matches the 
	algebra of the Poisson bracket. However, commutators 
	automatically satisfy the Jacobi identity, so it is
	impossible to match this algebra. Transforming to 
	canonical coordinates first, then quantizing, is impossible
	since canonical coordinates do not exist. 
%	We can't build a wave equation from the Hamilton-Jacobi
%	equation because \textcolor{red}{Why?} 
	Even path integral quantization is impossible because
	there is no Lagrangian action principle \cite{Rohr66,God82}.

	If quantization without the Jacobi identity is so difficult,
	how did Dirac quantize magnetic monopoles
	\cite{Dirac31,Dirac48}?
	
	Dirac proceeds by locally transforming to canonical 
	coordinates - by locally replacing fields with potentials.
	This comes at a cost. Dirac's theory has
	strings along which the electrons' wavefunctions are zero.
%	Dirac claims that this is not a physical constraint since
%	the locations of the strings are arbitrary. However, 
	Although the directions of the strings are arbitrary,
	the locations of the ends of the strings are not arbitrary
	because these are the locations of the monopoles.
	If we consider a monopole wave packet instead of a point 
	monopole, the electron must avoid each volume element of 
	the wave packet	\cite{Wen66}. 
%	By assuming that the electrons'
%	wavefunctions must be zero at the endpoints of the strings,
	Dirac implicitly assumes that electrons' and monopoles'
	positions could never coincide.
	This claim needs to be justified. If it were 
	true, it could restore the Hamiltonian nature of 
	electromagnetism %(See \S \ref{sec:1E1M}) 
	and remove the impediment to quantization.
	
	New quantization techniques are needed to build a 
	quantum field of magnetic monopoles. 	
	When expressed in terms of gauge group operators, 
	a violation of the Jacobi identity corresponds to a 
	nonzero 3-cocycle, which removes associativity of the 
	operators \cite{Lip69,Boul85,Jack85,WuZee85}.
	A nonassociative star 
	product for Wigner functions, % \cite{Sz1}, 
	doubling the size of the phase space to create a
	Hamiltonian structure on the extended space, % \cite{Sz2},
	and the geometric structure of a 
	gerbe % \cite{Sz3}
	have been used to address this \cite{SzRev}.
	Another possible tool is beatification, which 
	removes the explicit variable dependence from a
	Poisson bracket \cite{MorVan16}.
	
	Electromagnetism is often considered to be a long wavelength
	limit of a more fundamental theory with a broken gauge
	symmetry \cite{tHo74,Poly74}. 
	Magnetic monopoles appear when the symmetry is 
	broken in a topologically nontrivial way. In future
	work, we hope to write a classical SU(2) theory
	using an explicitly gauge invariant noncanonical
	Poisson bracket and check if it satisfies the Jacobi 
	identity. We would then break the gauge symmetry in a
	topologically nontrivial way to determine which aspects 
	of symmetry breaking are inherently quantum and which 
	are inherited from the classical theory.

\section{Conclusion}
\label{sec:Con}

	The existence of magnetic monopoles disrupts the 
	Hamiltonian nature of classical electromagnetism. When 
	the locations of an electron and a monopole coincide,
	the Jacobi identity for the Poisson bracket is violated.
	This result is most obviously seen in plasma physics,
	where the huge number of particles interacting collectively
	makes collisions between species almost guaranteed.	
	
	There are two ways to recover the Jacobi identity, but 
	neither is satisfactory. All species could 
	have the same ratio of magnetic to electric charge.
	This is a duality transformation away from the 
	universe we currently observe without monopoles. 
	Alternatively, we could insist that electrons' and 
	monopoles' positions never coincide, as Dirac's theory 
	implicitly assumed. 
	Why can't monopoles collide with ordinary matter? 
	How would this influence our attempts to detect them?
	
	Traditional methods of quantization fail without 
	the Jacobi identity, canonical coordinates, or
	a Lagrangian action principle.
	How should we quantize the interactions between 
	arbitrary collections of electrically and magnetically
	charged particles?
	
	Problems with the quantum theory of 
	electromagnetism with magnetic monopoles have been
	known for decades \cite{Lip69}. 
	In this letter, we showed that these
	problems are not inherently quantum. The quantum theory
	merely inherits the problems caused by the failure of 
	the Jacobi identity for the classical Poisson bracket.
	
	Since there is no experimental evidence, 
	the argument for magnetic monopoles is aesthetic. 
	The failure of the Jacobi identity taints this beauty.
	We should remain skeptical of any theory of magnetic
	monopoles that does not address the failure of the
	Jacobi identity. \\
	
	\noindent \textbf{Note:} 
	It has recently been shown that the bracket
	\refeq{VMPBM} not only violates the Jacobi identity, 
	it also does not satisfy the weaker conditions for a twisted
	Poisson bracket \cite{LaiSarWei19}.
	
	\section*{Acknowledgment}
	\noindent   
	We would like to thank Behzat Ergun, Robert Szabo,
	and Michael Waterbury for guidance in
	navigating the high energy theory literature.
	Supported by U.S. Dept.\ of Energy Contract \# DE-FG05-80ET-53088. 
	
\bibliographystyle{elsarticle-num}

\bibliography{mono}

\begin{thebibliography}{10}
\expandafter\ifx\csname url\endcsname\relax
  \def\url#1{\texttt{#1}}\fi
\expandafter\ifx\csname urlprefix\endcsname\relax\def\urlprefix{URL }\fi
\expandafter\ifx\csname href\endcsname\relax
  \def\href#1#2{#2} \def\path#1{#1}\fi

\bibitem{Dirac31}
P.~A.~M. Dirac, Quantised singularities in the electromagnetic field, Proc.
  Roy. Soc. A 133 (1931) 60--72.

\bibitem{Dirac48}
P.~A.~M. Dirac, The theory of magnetic poles, Phys. Rev. 74~(7) (1948)
  817--830.

\bibitem{WuYa76}
T.~T. Wu, C.~N. Yang, Dirac's monopole without strings: Classical lagrangian
  theory, Phys. Rev.D 14~(2) (1976) 437--445.

\bibitem{Bir71}
I.~Bialynicki-Birula, Z.~Bialynicki-Birula, Magnetic monopoles in the
  hydrodynamic formation of quantum mechanics, Phys. Rev. D 3~(10) (1971)
  2410--2412.

\bibitem{CabFer62}
N.~Cabibbo, E.~Ferrari, Quantum electrodynamics with dirac monopoles, Nuovo
  Cimento 23~(6) (1962) 1147--1154.

\bibitem{Schw66}
J.~Schwinger, Magnetic charge and quantum field theory, Phys. Rev. 144~(4)
  (1966) 1087--1093.

\bibitem{Schw68}
J.~Schwinger, Sources and magnetic charge, Phys. Rev. 173~(5) (1968)
  1536--1544.

\bibitem{Schw69}
J.~Schwinger, A magnetic model of matter, Science 165~(3895) (1969) 757--760.

\bibitem{tHo74}
G.~'t~Hooft, Magnetic monopoles in unified gauge theories, Nuc. Phys. B 79
  (1974) 276--284.

\bibitem{Poly74}
A.~M. Polyakov, Particle spectrum in the quantum field theory, JETP Letters 20
  (1974) 194--195.

\bibitem{Pol04}
J.~Polchinski, Monopoles, duality, and string theory, Int. J. Mod. Phys. A
  19~(supp01) (2004) 145--154.

\bibitem{Kib76}
T.~Kibble, Topology of cosmic domains and strings, J. Phys. A 9~(8) (1976)
  1387--1398.

\bibitem{Pres79}
J.~P. Preskill, Cosmological production of superheavy magnetic monopoles, Phys.
  Rev. Lett. 43~(19) (1979) 1365--1368.

\bibitem{Guth81}
A.~H. Guth, Inflationary universe: A possible solution to the horizon and
  flatness problems, Phys. Rev. D 23~(2) (1981) 347--356.

\bibitem{Par70}
E.~N. Parker, The origin of magnetic fields, Ap. J. 160 (1970) 383--404.

\bibitem{Pri75}
P.~B. Price, E.~K. Shirk, W.~Z. Osborne, L.~S. Pinsky, Evidence for detection
  of a moving magnetic monopole, Phys. Rev. Lett. 35~(8) (1975) 487--490.

\bibitem{Cab82}
B.~Cabrera, First results from a superconductive detector for moving magnetic
  monopoles, Phys. Rev. Lett. 48~(20) (1982) 1378--1381.

\bibitem{Alv75}
L.~W. Alvarez, Analysis of a reported magnetic monopole (1975).

\bibitem{Pat16}
C.~Patrignani, et~al. (Particle Data~Group), Review of particle physics,
  Chinese Phys. C 40~(100001) (2016) 1675--1682.

\bibitem{Mil06}
K.~A. Milton, Theoretical and experimental status of magnetic monopoles, Rep.
  Prog. Phys. 69 (2006) 1637--1711.

\bibitem{Raj12}
A.~Rajantie, Magnetic monopoles in field theory and cosmology, Phil. Trans.
  Roy. Soc. A 370 (2012) 5705--5717.

\bibitem{Atiyah}
M.~F. Atiyah, N.~Hitchin, The geometry and dynamics of magnetic monopoles,
  Princeton University Press, 1988.

\bibitem{Shnir}
Y.~M. Shnir, Magnetic monopoles, Springer, 2005.

\bibitem{Born1}
M.~Born, L.~Infeld, Foundations of the new field theory, Proc. Roy. Soc. A 144
  (1934) 425--451.

\bibitem{Born2}
M.~Born, L.~Infeld, On the quantization of the new field equations., Proc. Roy.
  Soc. A 147 (1934) 522--546.

\bibitem{pjm80}
P.~J. Morrison, The {M}axwell-{V}lasov equations as a continuous {H}amiltonian
  system 80 (1980) 383--386.

\bibitem{pjm82}
P.~J. Morrison, Poisson brackets for fluids and plasmas, AIP Conf. Proc. 88
  (1982) 13--46.

\bibitem{MW82}
J.~E. Marsden, A.~Weinstein, The hamiltonian structure of the maxwell-vlasov
  equations, Physica D 4~(3).

\bibitem{Bir84}
I.~Bialynicki-Birula, J.~C. Hubbard, L.~A. Turski, Gauge-independent canonical
  formulation of relativistic plasma theory, Physica A 128~(3) (1984) 509.

\bibitem{Morr98}
P.~J. Morrison, {H}amiltonian description of the ideal fluid 70 (1998)
  467--521.

\bibitem{Morr13}
P.~J. Morrison, A general theory for gauge-free lifting, Phys. Plasmas 20
  (2013) 012104.

\bibitem{Jackson}
J.~D. Jackson, Classical Electrodynamics, 3rd Edition, Wiley, 1998, pp.
  273--275.

\bibitem{sudarshan}
E.~Sudarshan, N.~Mukunda, Classical Dynamics: A Modern Perspective, Wiley, New
  York, 1974.

\bibitem{ArKozNei}
V.~I. Arnold, V.~V. Kozlov, A.~I. Neishtadt, Mathematical Aspects of Classical
  and Celestial Mechanics, 3rd Edition, Springer, 2006.

\bibitem{souriau}
J.-M. Souriau, Structure des syst{\`e}mes dynamiques, Dunod, Paris, 1970.

\bibitem{thirring}
W.~Thirring, Classical Mathematical Physics: Dynamical Systems and Field
  Theories, 3rd Edition, Springer Verlag, New York, 1997.

\bibitem{DA15}
E.~C. D'Avignon, Physical consequences of the jacobi identity, ArXiv
  e-prints\href {http://arxiv.org/abs/1510.06455} {\path{arXiv:1510.06455}}.

\bibitem{LJ82}
R.~G. Littlejohn, Singular poisson tensors, AIP Conf. Proc. 88 (1982) 47--66.

\bibitem{Eis61}
L.~P. Eisenhart, Continuous Groups of Transformations, 2nd Edition, Dover,
  1961, p. 250.

\bibitem{Rohr66}
F.~Rohrlich, Classical theory of magnetic monopoles, Phys. Rev. 150~(4) (1966)
  1104--1111.

\bibitem{God82}
J.~Godfrey, Do dirac monopoles admit a lagrangian?, Nuovo Cimento 71 A~(1)
  (1982) 134--146.

\bibitem{Wen66}
G.~Wentzel, Comments on dirac's theory of magnetic monopoles, Supp. Prog.
  Theor. Phys. 37 (1966) 163--174.

\bibitem{Lip69}
H.~J. Lipkin, W.~I. Weisberger, M.~Peshkin, Magnetic charge quantization and
  angular momentum, Ann. Phys. 53 (1969) 203--214.

\bibitem{Boul85}
D.~G. Boulware, S.~Deser, B.~Zumino, Absence of 2-cocycles in the dirac
  monopole problem, Phys. Lett. 153B~(4,5) (1985) 307--310.

\bibitem{Jack85}
R.~Jackiw, Three-cocycle in mathematics and physics, Phys. Rev. Lett. 54~(3)
  (1985) 159--162.

\bibitem{WuZee85}
Y.-S. Wu, A.~Zee, Cocycles and magnetic monopole, Phys. Lett. 152~(1,2) (1985)
  98--102.

\bibitem{SzRev}
R.~J. Szabo, Higher quantum geometry and non-geometric string theory, in: 17th
  Hellenic School and Workshops on Elementary Particle Physics and Gravity,
  Corfu, Greece, 2018.
\newblock \href {http://arxiv.org/abs/1803.08861} {\path{arXiv:1803.08861}}.

\bibitem{MorVan16}
P.~J. Morrison, J.~Vanneste, Weakly nonlinear dynamics in noncanonical
  hamiltonian systems with applications to fluids and plasmas, Ann. Phys. 368
  (2016) 117--147.

\bibitem{LaiSarWei19}
M.~Lainz, C.~Sard{\'o}n, A.~Weinstein, Plasma in a monopole background is not
  twisted poisson, To appear in Phys. Rev. D\href
  {http://arxiv.org/abs/1908.03986} {\path{arXiv:1908.03986}}.

\end{thebibliography}

\end{document}